\documentclass[sigconf]{acmart}
\makeatletter
\renewcommand\@formatdoi[1]{\ignorespaces}
\makeatother
\renewcommand\footnotetextcopyrightpermission[1]{} 
\AtBeginDocument{%
  \providecommand\BibTeX{{%
    \normalfont B\kern-0.5em{\scshape i\kern-0.25em b}\kern-0.8em\TeX}}}

\copyrightyear{2021}
\acmYear{2021}
\setcopyright{rightsretained}
\acmConference[]{}{}{}
\acmBooktitle{CHI Conference on Human Factors in Computing Systems Extended Abstracts (CHI '21 Extended Abstracts), May 8--13, 2021, Yokohama, Japan}\acmDOI{10.1145/3411763.3443436}
\acmISBN{978-1-4503-8095-9/21/05}

\begin{document}

\title[]{Designing an Interactive Visualization System for Monitoring Participant Compliance in a Large-Scale, Longitudinal Study}


\author{Poorna Talkad Sukumar}
\affiliation{%
  \institution{University of Notre Dame}}
\email{ptalkads@nd.edu}

\author{Thomas Breideband}
\affiliation{%
  \institution{University of California, Irvine}
}
\email{tbreideband@gmail.com}

\author{Gonzalo Martinez}
\affiliation{%
 \institution{University of Notre Dame}}
 \email{gmarti11@nd.edu}

\author{Megan Caruso}
\affiliation{%
  \institution{University of Colorado Boulder}}
  \email{megan.caruso@colorado.edu}
  
\author{Sierra Rose}
\affiliation{%
  \institution{University of California, San Diego}}
\email{snrose@ucsd.edu}

\author{Cooper Steputis}
\affiliation{%
  \institution{University of Colorado Boulder}}
\email{cost5872@colorado.edu}

\author{Sidney D'Mello}
\affiliation{%
  \institution{University of Colorado Boulder}}
\email{sidney.dmello@gmail.com}

\author{Gloria Mark}
\affiliation{%
  \institution{University of California, Irvine}}
\email{gmark@uci.edu}

\author{Aaron Striegel}
\affiliation{%
  \institution{University of Notre Dame}}
\email{striegel@nd.edu}

\renewcommand{\shortauthors}{}

\begin{abstract}

Frequent monitoring of participant compliance is necessary when conducting large-scale, longitudinal studies to ensure that the collected data is of sufficiently high quality. While the need for achieving high compliance has been underscored and there are discussions on incentives and factors affecting compliance, little is shared about the actual processes and tools used for monitoring compliance in such studies.  
Monitoring participant compliance with respect to multi-modal data can be a tedious process, especially if there are only a few personnel involved. In this case study, we describe the iterative design of an interactive visualization system we developed for monitoring compliance and refined based on changing requirements in an ongoing study. We find that the visualization system, leveraging the digital medium, both facilitates the exploratory tasks of monitoring participant compliance and supports asynchronous collaboration among non-co-located researchers. Our documented requirements for checking participant compliance as well as the design of the visualization system can help inform the compliance-monitoring process in future studies.
\end{abstract}



\maketitle
\thispagestyle{empty}

\begin{figure*}[h]
  \centering
  \includegraphics[width=0.85\linewidth]{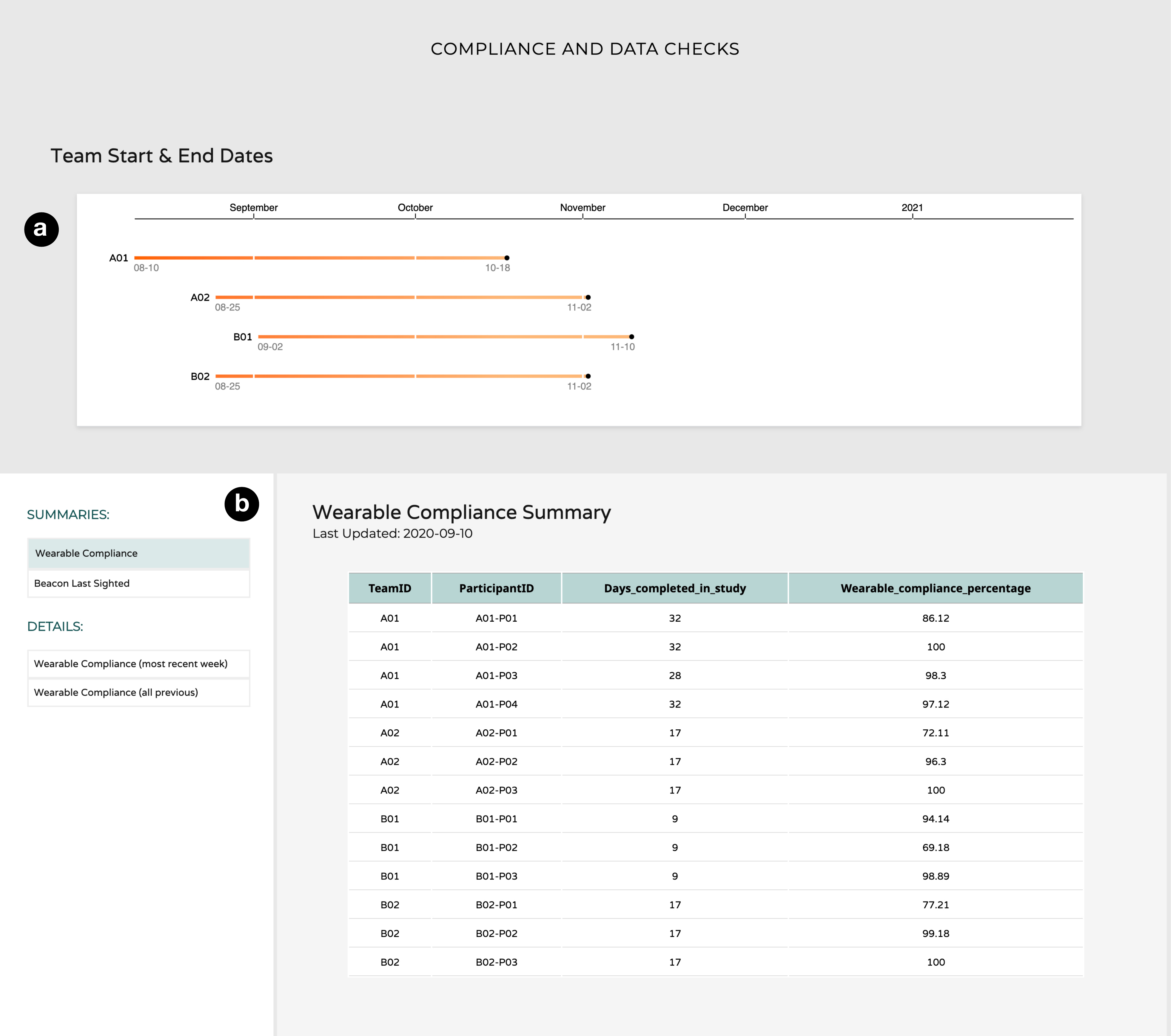}
  \caption{A screen capture of the initial design of our visualization interface for monitoring compliance. (a) Study context provided by a static timeline chart showing the start and end dates of teams currently enrolled in the study. (b) A set of tables presenting compliance and beacon-monitoring information of participants. (Figure shows synthetic data). }
  \label{initial_design}
\end{figure*}

\section{Introduction}

Large-scale, longitudinal studies, gathering diverse types of data (such as, phone usage, physical activity, and affect) and requiring little active engagement by participants, provide an unprecedented means to gain insights into individual and collective human behavior. Examples of such studies include \textit{StudentLife} \cite{wang2014studentlife}, where various types of behavioral data were collected from 48 college students over a 10-week period through a smartphone app; and \textit{Tesserae} \cite{mattingly2019tesserae}, where various personal attributes of 757 information workers across five organizations were tracked for a year through wearables and Bluetooth beacons. 
The implementations of such studies generally include mechanisms for monitoring compliance wherein the (multi-modal) data being collected from each participant is frequently checked to identify and address issues of missing data. While studies typically report and discuss achieved compliance retrospectively \cite{faust2017exploring, wang2014studentlife, striegel2013lessons, harrison2014tracking, mattingly2019tesserae, mundnich2020tiles}, they provide little information about how compliance was monitored during the study and the tools that were used for the process.

In this case study, we focus solely on the compliance-monitoring process in an ongoing, large-scale, tracking study \cite{futureofwork}, where we are gathering various health, activity, proximity, and work-related information from \emph{teams} for a period of ten weeks, using a staggered enrollment process. From our prior experiences with conducting similar studies \cite{mattingly2019tesserae, striegel2013lessons, purta2016experiences}, we find that the process of checking compliance (and subsequent nudging of participants) cannot be fully automated. The process requires a human in the loop to explore details of the computed compliance for each participant 
and make careful judgments 
drawing from an understanding of how the devices work and participant data from the devices is gathered in our backend database.

We designed an interactive, web-based visualization system for monitoring the compliance of participants in our study and refined the design based on our growing requirements as the study progressed. Our experiences with this system to date have revealed that its advantage is two-fold. First, the system facilitates monitoring compliance. This is because compliance monitoring is essentially a data-exploration task and interactive visualization systems are best suited to support exploratory tasks \cite{fekete2008value}. Second, by being web-based, the system supports asynchronous collaboration among non-co-located researchers. The content of our interface, consisting of a collection of CSV files, can be updated by any of the non-co-located researchers in the study who have the necessary permissions. 

Our aim in this paper is to outline our requirements for monitoring compliance and describe the iterative design of our visualization system and the rationales behind the decisions we made. We also discuss further refining the system and implementing additional features which could be beneficial. Our paper can help inform the compliance-monitoring process in future studies and makes a case for the use of visualizations for this process.

\section{Background}

We describe our study design and the process of monitoring participant compliance.

\subsection{Our Study Design}

Our ongoing, multi-university collaborative study \cite{futureofwork} is concerned with studying \textit{teamwork}, including aspects relating to team dynamics, diversity, and performance, by collecting data from three-to-five person teams in complex information work professions for a period of ten weeks. This work is funded by two grants (a factor considered in the design of our visualization interface) and each grant covers expenses corresponding to roughly half of the teams enrolled in the study. Our goal is to collect data from 70 teams and we have enrolled 15 teams thus far.

Participants are each provided with a Garmin vívosmart 4 wearable, which they are required to wear at all times during the study (except when charging the wearable). Various health and activity metrics, such as, sleep, steps, heart rate, and stress, are gathered from the wearable. Participants are also required to place a Gimbal Series 21 Bluetooth beacon in their workspaces and install an app on their phones. The app captures sightings of the beacon whenever the phone comes into proximity of the beacon. In addition to these unobtrusive sensing streams, we also use the experience sampling method (ESM) to gather participant responses to short questionnaires sent every weekday via a proprietary mobile survey app \cite{expiwell}.

\subsection{Compliance Monitoring in Longitudinal Studies} 

In large-scale, longitudinal studies, it is crucial to frequently monitor participant compliance to minimize the occurrence of missing data. While prior studies discuss the challenges of achieving high participant compliance \cite{purta2016experiences, faust2017exploring, striegel2013lessons, mattingly2019tesserae, wang2014studentlife}, they do not discuss the challenges of the very process of monitoring compliance. Monitoring compliance, especially with respect to sensing streams, can be non-trivial. The process is not as straightforward as identifying participants with missing data and nudging them to be more compliant. Low compliance or missing data corresponding to a participant can signal many things, including participant non-compliance, issues with participant's devices, data syncing delays, patterns in device usage (e.g., device not worn at night), and more significant technical issues affecting all or a big subset of the participants \cite{purta2016experiences, stopczynski2014measuring, mattingly2019tesserae, harrison2014tracking, striegel2013lessons}. We ourselves have encountered all of these situations in our study so far. Therefore, while compliance scores can be automatically generated, these scores require reasoned interpretations and actions by study personnel.  For example, if we observe missing data for participants for most recent dates, these are most likely the result of data syncing delays. We also need to consider any issues reported by participants to account for their provisional low compliance.  Therefore, we regard monitoring compliance as an analytic or data-exploration task involving careful consideration of participant compliance data and one that is best supported by visualization tools \cite{fekete2008value}. 

In our study, we use the heart rate (HR) data from participant wearable, which is updated every 15 seconds, to calculate wearable compliance. For each participant, we check if HR data is present in every 48 non-overlapping half-hour windows in a day and assign a score of 1 (present) or 0 (not present) accordingly. The daily compliance is then computed as (sum of scores for the day)/48*100. 
Computing survey compliance is straightforward--daily compliance is 100\% if survey is completed by the participant and 0\% otherwise. Participants are encouraged to maintain average daily wearable and survey compliance scores of at least 80\%. We monitor only the ``last sighting'' information for each participant beacon and if a beacon has not been sighted for more than a few days, we contact the participant to make sure that both the beacon and the beacon app are working correctly.

\begin{figure*}[t]
  \centering
  \includegraphics[width=0.96\linewidth]{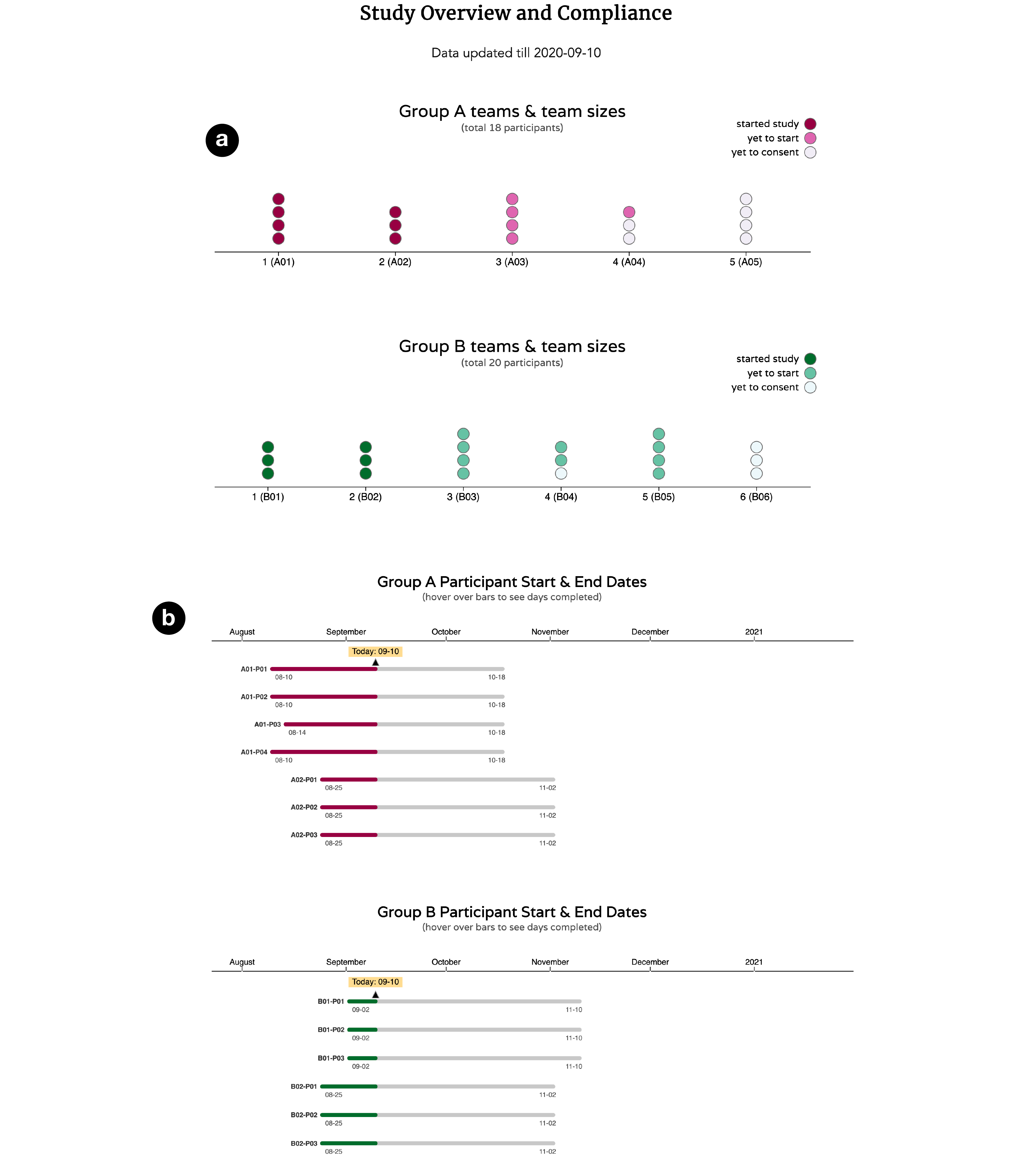}
  \caption{A screen capture of the top portion of our improved visualization interface presenting study context. (a) Dot-array-based visualizations presenting number of teams, team sizes, and study status of participants for each funding source (Group A and Group B). (b) Timeline charts presenting participant timelines in the study for each funding source. (Figure shows synthetic data).}
  \label{refined_design1}
\end{figure*}

\section{Initial Design of the Visualization System}

We describe the initial design of our visualization system for monitoring compliance including the data visualized, the abstract visualization tasks of the users supported by the system, the visual representations chosen for the data, and the technical implementation.

\subsection{Data}
\label{data}

For each participant currently in the study, we required the following data for monitoring their compliance:

\begin{enumerate}
    \item \textbf{\textit{Identifiers:}} assigned team and participant IDs, (which are non-personal identifiers), 
    \item \textbf{\textit{Context:}} team start and end dates and number of days completed in the study,
    \item \textbf{\textit{Wearable compliance:}} compliance (\%) for wearable data with three levels of granularity (half-hourly, daily, and overall) 
    \item \textbf{\textit{Survey compliance:}} compliance (\%) for surveys with two levels of granularity (daily and overall) 
    \item \textbf{\textit{Beacon-monitoring information:}} date of the last sighting of participant's beacon.
    
\end{enumerate}

\subsection{Tasks}

There are two categories of users within our research group, each with distinct tasks and interests in the data. We present the abstract visualization tasks of these user categories below.
\begin{enumerate}
    \item \textbf{\textit{Category I:}} the principal investigators (PIs) of the projects are mainly interested in the overview and mostly \textbf{browse} the overall compliance of the participants. 
    \textit{Browse} refers to the search action performed when a user has no particular target in mind but knows the location of the information they are interested in \cite{munzner2014visualization}(Ch. 3).
    \item \textbf{\textit{Category II:}} the graduate students and postdoctoral researcher are concerned with the details and they \textbf{lookup} each participant and \textbf{identify} the participant's daily wearable and survey compliance and beacon-monitoring information. 
    \textit{Lookup} refers to the search action performed when a user both  has a particular target in mind and knows its location; \textit{identify} is a query action performed to retrieve values corresponding to a single target (in this case, a participant) returned by the \textit{lookup} action \cite{munzner2014visualization}(Ch. 3). 
    They also \textbf{browse} compliance and beacon-monitoring information of participants and \textbf{compare} compliance scores both belonging to the same participant and across participants.
\end{enumerate}

\subsection{Visual Representations}

The context information consisting of team start and end dates are visualized using a static timeline chart in our interface. We visualize the compliance and beacon-monitoring data using tables. Tables are a simple and effective tool for visualizing data \cite{wilke2019fundamentals}(Ch. 22) and they support all of the aforementioned user tasks. They can, however, become unwieldy as the data grows and may require additional interaction mechanisms to support queries as we describe in Section \ref{improvements}.  A screen capture of our initial design of the visualization interface with the timelines and tables is presented in Fig. \ref{initial_design}. 

To facilitate user tasks, we present a set of different tables in the interface and users can select to view any of these tables. For example, one of the tables presents the wearable compliance summary of participants to facilitate tasks of users in category I. The users in category II conduct compliance checks every week where they look at the daily compliance scores corresponding to the \textit{most recent week} for each participant. In some cases, they also look at \textit{all previous} daily compliance of a participant to make judgments on whether or not to nudge them. Hence we also created separate tables for these two sets of data (``most recent week'' and ``all previous''). 

While the \textit{half-hourly} wearable compliance data can be useful to discover patterns in wearable usage (for example, wearables not being worn during the night), we observed that this data was almost never considered and daily compliance is the preferred granularity level for conducting the compliance checks. Hence we did not include the half-hourly compliance data in our interface. Furthermore, since we are using proprietary software \cite{expiwell} for gathering survey responses in our study, survey compliance was initially computed manually and maintained separately and not included in the interface.

\begin{figure*}[t]
  \centering
  \includegraphics[width=0.85\linewidth]{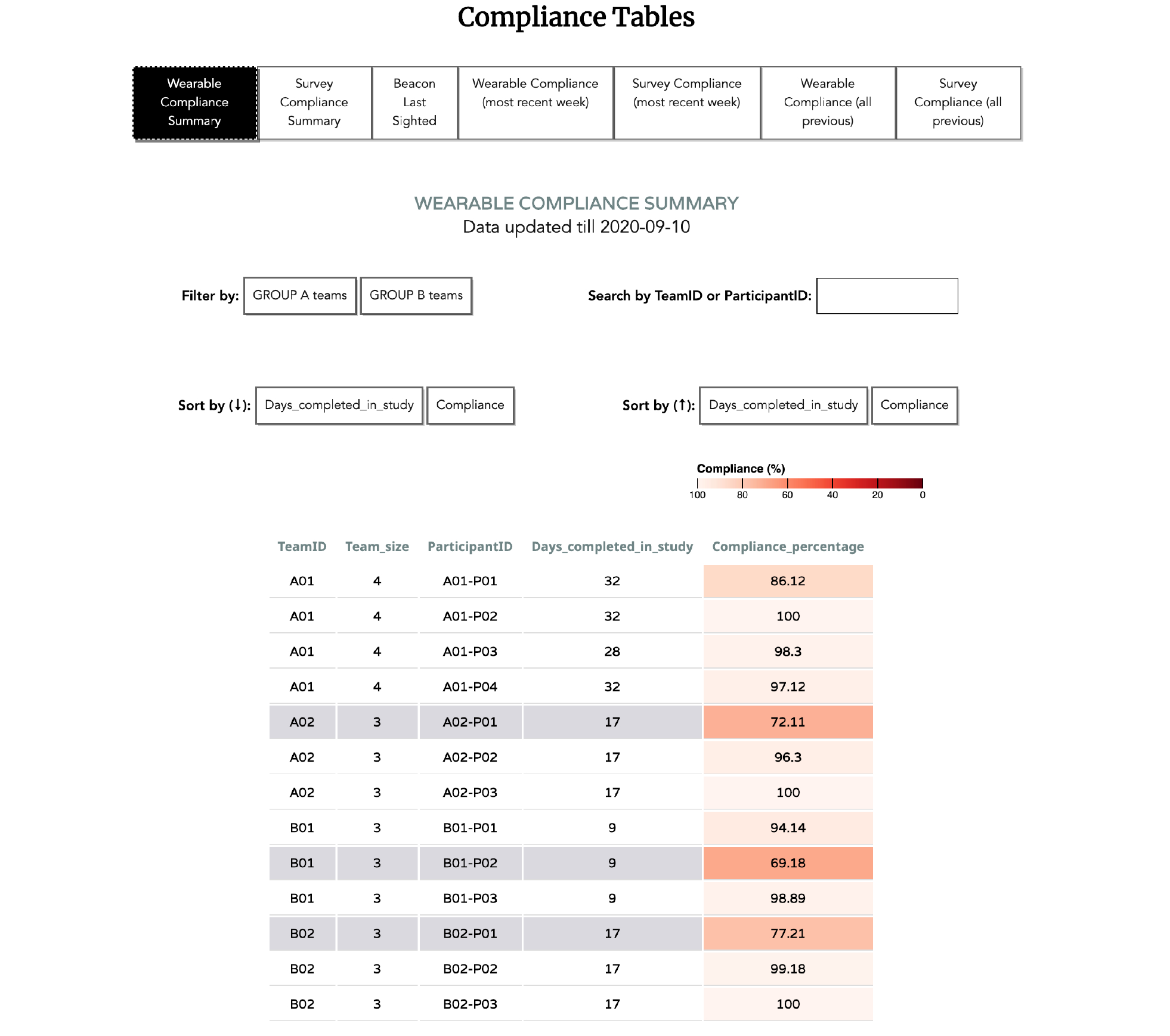}
  \caption{A screen capture of the bottom portion our improved visualization interface containing the compliance tables. Improvements made include addition of survey-compliance tables, options to sort and filter table contents, multiple-row selection feature, and visual encoding of compliance scores. (Figure shows synthetic data).}
  \label{refined_design2}
\end{figure*}

 \begin{figure*}[t]
  \centering
  \includegraphics[width=0.85\linewidth]{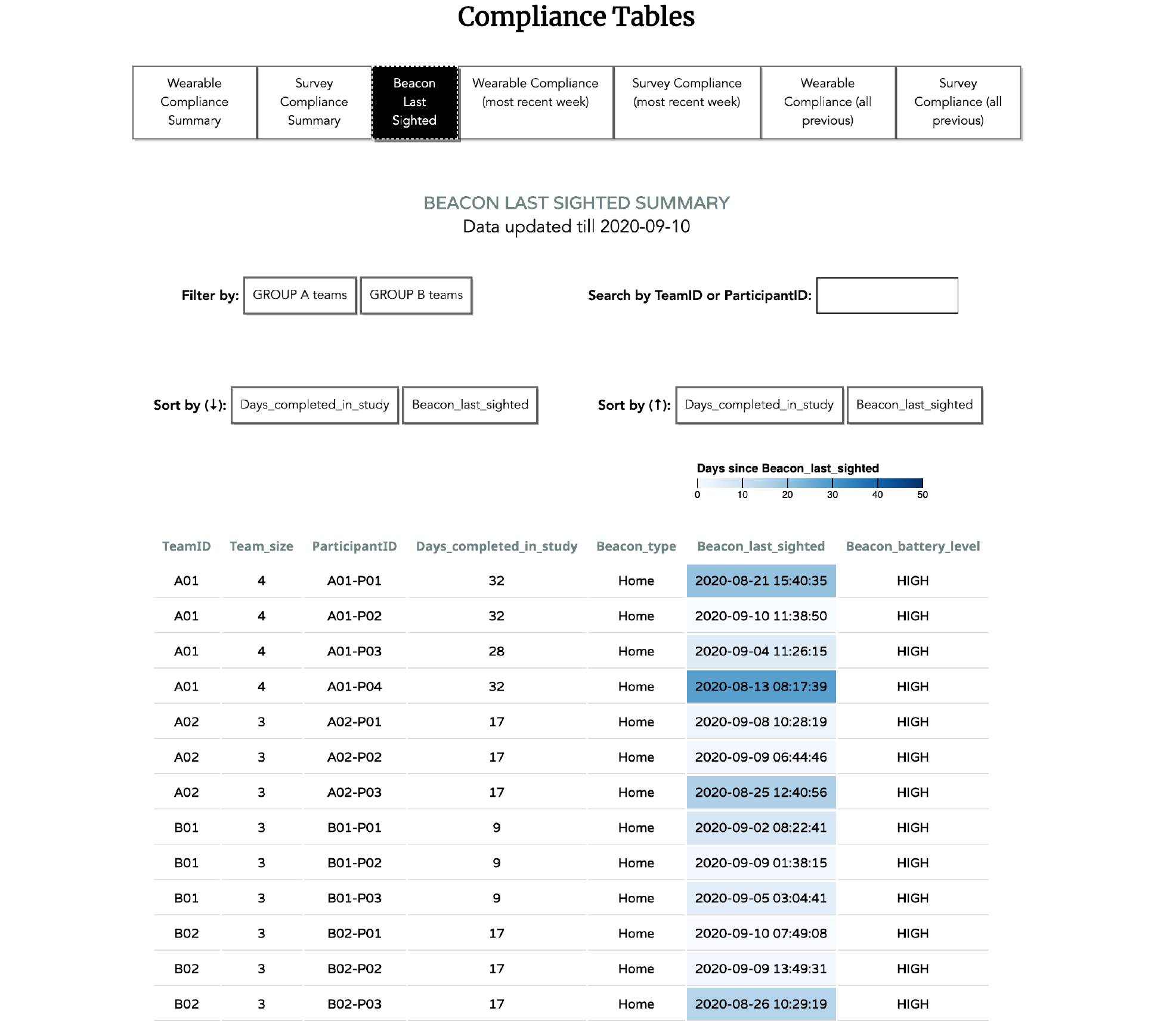}
  \caption{Same as Fig. \ref{refined_design2} but with the "Beacon Last Sighted" option selected, showing the visual encoding of the ``days since last beacon sighting'' information. (Figure shows synthetic data).}
  \label{refined_design3}
\end{figure*}



\subsection{Technical Implementation}

The web-based visualization interface was developed using D3.js \cite{D3.js}.   The data for the visualizations consist of a collection of CSV files. These files are updated 1-2 times every week via SFTP (Secure File Transfer Protocol) by running Python scripts that query both the backend database (to compute participant wearable compliance and retrieve beacon sightings) as well as an enrollment spreadsheet containing study details. 

Our technical implementation is designed to support asynchronous collaboration among the non-co-located researchers in our study and to facilitate task hand-offs. Any of the researchers in the study (with the necessary permissions to access our web server) can update the CSV files. The visualization interface is also password-protected and can only be viewed by researchers in the study. Furthermore, the interface and the CSV files contain no personally identifiable information. 

\section{Improvements made to the Visualization System}
\label{improvements}

While our initial, largely-static design worked well with a small number of participants, as more number of participants were enrolled in the study, it became increasingly challenging to perform our tasks using the interface. Hence, we made many improvements to our interface, drawing from visualization design principles and guidelines, to handle the visual complexity and to satisfy our growing requirements. These improvements were made over the course of two months and are outlined below in the same order in which they were made. Screen captures of our improved visualization system are presented in Figures \ref{refined_design1}, \ref{refined_design2}, and \ref{refined_design3}.

\begin{enumerate}
    \item \textbf{\textit{Visualizing participant (instead of team) timelines: }} We found the team timelines in the initial design (see Fig. \ref{initial_design}(a)) to be too high-level and potentially misleading in cases where there were one or two participants within a team with delayed start dates. Hence, we changed the timeline chart to represent participant timelines instead of team timelines.

    \vspace{0.1in}
    \item \textbf{\textit{Presenting additional study context: }} The users in Category I wanted to see more study-related information on the interface including the number of teams currently enrolled in the study, the number of participants within each team, and teams in the enrollment pipeline. Since our study has two different funding sources, they also wanted to see and compare the teams under each funding source. 

We use a dot-array-based visualization \cite{chartchooser} to present this additional study information as shown in Fig. \ref{refined_design1}(a), where each vertical array of dots represents one team and the number of dots in an array represent the participants within the team. We also visually encode the study status of each participant (``started study'', ``yet to start'', and ``yet to consent'') using sequential color scales. The dot-array visualizations corresponding to each funding source, labelled ``Group A'' and ``Group B'' in Fig. \ref{refined_design1}(a), are arranged one below the other to enable comparisons between the two groups. 

 \vspace{0.1in}
 
\item \textbf{\textit{Visualizing participant progress: }} In addition to the start and end dates of the participants in the timeline chart, the users in Category I wanted to see the progress of each participant in the study. They also wanted the participants to be separated by funding source. 
We improved the static timeline chart to also show the progress of participants in the study as shown in Fig. \ref{refined_design1}(b). Users can also hover over each bar in the chart to see the number of days completed by the participant in the study. 

 \vspace{0.1in}
 
\item \textbf{\textit{Dealing with visual complexity in the tables: }} With an increase in the number of participants in our study (resulting in more rows in the tables), it became difficult for users in both categories to perform their respective tasks using the interface. Hence we implemented additional interaction mechanisms, commonly found in the visualization literature \cite{munzner2014visualization}(Ch. 11-13), to deal with the visual clutter in the tables.  

We implemented options to \textit{reorder/sort} attributes (such as, compliance \%, and date) in the tables. We also included mechanisms for \textit{reducing} rows by implementing options to search (by participant ID or team ID) and filter teams (by Group A or Group B).  
Users could also select/highlight multiple rows at the same time, a feature that was especially useful to users in Category II for keeping track of participants to nudge when doing compliance checks. The added interaction options can be seen in Fig. \ref{refined_design2}.

 \vspace{0.1in}
 
\item \textbf{\textit{Adding survey compliance: }} As mentioned previously, we initially maintained survey compliance separately but we found it tedious to conduct separate compliance checks for surveys and preferred a unified system presenting compliance scores for all the different data streams. Hence we also included survey compliance in our visualization system by writing additional scripts to compute survey compliance from the survey responses. Fig. \ref{refined_design2} shows the added options for viewing the survey-compliance tables.

 \vspace{0.1in}
 
\item \textbf{\textit{Color coding the main attributes: }} As the number of participants in the study increased, the users in Category II wanted to be able to quickly browse and identify participants with low compliance scores in the tables when performing compliance checks. To enable this task, we encoded the compliance-score cells using a sequential color scale. 
Similarly, we also encoded the ``days since last beacon sighting'' information using a sequential color scale to enable quick identification of participants with potential beacon-related issues.
These additional encodings for compliance and beacon sightings can be seen in Figures \ref{refined_design2} and \ref{refined_design3}, respectively. 

 \vspace{0.1in}
 
\item \textbf{\textit{Responsive design: }} Our interface can be viewed on any device with screen width greater than 740 px (includes all iPads and most tablet devices) and only the visualizations shown in Fig. \ref{refined_design1} are displayed on mobile devices with screen width less than 740 px. We, however, find that the exploratory tasks of users in Category II are more conveniently performed on large displays.

\end{enumerate}

\section{Discussion and Future Work} 

Based on our experiences, we present implications for the design of visualization systems for monitoring compliance and also discuss further refining our visualization system.

\begin{enumerate}
    \item \textbf{\textit{Effectiveness of interactive tables for monitoring compliance: }} While tabular visualizations have received attention in the context of decision-making \cite{dimara2016attraction, dimara2017conceptual} and data-wrangling tasks \cite{kasica2020table} 
    within the visualization literature, from our experiences, we find that they also facilitate the data-exploration task of monitoring compliance, provided interaction mechanisms are implemented to support querying the table contents. 
    A potential refinement would be to support \textit{linked highlighting} \cite{munzner2014visualization}(Ch. 12) between the different tables included in the interface where participants interactively selected in one table are immediately highlighted in all the other tables. Similarly, linked highlighting between the tables, timelines, and dot-array visualizations could also be useful where participants selected in one view are immediately highlighted in all the other views.
    
    \vspace{0.07in}

    \item \textbf{\textit{Supporting requirements of different categories of users: }}Similar to our case, compliance data may be of interest to different categories of users and it is important to design interfaces that support the tasks of these different user categories. Participants could also be potential stakeholders of the system and it may be useful to implement additional functionality that enables each participant to monitor their own compliance with respect to the different data streams and also report issues as done in prior studies \cite{purta2016experiences, mattingly2019tesserae}. 
    
    \vspace{0.07in}
    
    \item \textbf{\textit{Supporting asynchronous collaboration: }}While our interface supports content updating by non-co-located personnel, it only scratches the surface of what is possible in terms of asynchronous collaboration. Drawing from prior work \cite{viegas2007manyeyes, heer2007voyagers}, we can implement capabilities such as graphical annotation and adding comments containing snapshots of the data. Such capabilities can enable researchers in the study to communicate their compliance-related discoveries and judgments along with visual data evidence to others asynchronously.
    
    \vspace{0.07in}
    
    \item \textbf{\textit{Designing a unified system: }}We find that it is very convenient to perform compliance checks using a single system presenting compliance scores across all the data streams. Hence, in studies gathering multi-modal data, implementing compliance-monitoring systems consolidating compliance data from the different data streams can be beneficial.
    
    

\end{enumerate}

\section{Conclusions}

We highlight the challenges in monitoring participant compliance when conducting large-scale, longitudinal studies and describe a visualization system designed for this purpose in an ongoing study. We find that our visualization system, presenting compliance data in the form of \textit{interactive tables}, facilitates the exploratory task of checking participant compliance. Based on our experiences, we present implications for the design of such systems, including supporting requirements of different stakeholders and implementing functionality for asynchronous collaboration. 
This work commends the use of visualization tools in the process of compliance monitoring in longitudinal studies, where visualizations are already being used for other aspects, for example, to present participants with their respective data collected during the study \cite{cuttone2013mobile, s.20201052}.

\begin{acks}
This material is based upon work supported by the National Science Foundation under Grant No. SES-1928645, SES-1928718, SES-1928612, and SES-2030599.
\end{acks}

\bibliographystyle{ACM-Reference-Format}
\bibliography{sample-base}


\end{document}